\documentclass{article}

\usepackage{amssymb}
\usepackage{graphicx}
\usepackage{hyperref}
\usepackage{subfigure}
\usepackage{amsmath}
\usepackage{color}
\usepackage{authblk}

\title{Time-Dependent Ginzburg-Landau Simulations of Superconducting Vortices in Three Dimensions}
\author[1]{Antonio Lara}
\author[1]{C\'esar Gonz\'alez-Ruano}
\author[1]{Farkhad G. Aliev}
\affil[1]{Dpto. Fisica de la Materia Condensada C-III, Instituto Nicolas Cabrera (INC) and Condensed Matter Physics Institute (IFIMAC), Universidad Autonoma de Madrid, Madrid 28049, Spain}

\affil[*]{\textit{email: farkhad.aliev@uam.es}}
\date{\today }

\begin{document}

\maketitle

\begin{abstract}
Here we describe a development of computer algorithm to simulate the Time Dependent Ginzburg-Landau equation (TDGL) and its application to understand superconducting vortex dynamics in confined geometries. Our initial motivation to get involved in this task was trying to understand better our experimental measurements on dynamics of superconductors with vortices at high frequencies leading to microwave stimulated superconductivity due to presence of vortex (Lara, et al., Scientific Reports, 5  9187 (2015)).
\end{abstract}

\section{Time-dependent Ginzburg-Landau simulations}
High frequency dynamics in superconducting and magnetic vortex states has been subject of recent activities both from experimental and numerical points of view. While Landau Lifshitz Gilbert equation is simulated numerically to obtain static magnetization and spin wave modes \cite{Awad2010}, 
the Time Dependent Ginzburg-Landau equation is usually resolved numerically to understand the static and dynamic properties in the superconductong vortex state \cite{Lara2015}. 

The initial attempts to simulate the vortex behavior as a function of time using the TDGL equation date from the early 1990s. The most popular method to tackle this problem is the link variable method \cite{tdgl2, tdgl3, tdgl4, tdgl5, tdgl6, tdgl7}, which uses finite differences (i.e. regular grids and derivatives are approximated by subtractions between neighbors). Some attempts have been made using finite elements \cite{tdgl1}, but the extra complications that this method has makes it much harder to implement, although it allows to better represent curved shapes. 

The work presented below is based on the numerical approach exposed in \cite{Buscaglia}, with important extensions to include arbitrary shapes and calculations in 3D. First, some general concepts about numerical solution of differential equations are explained, followed by a more detailed description of the particular method for solving TDGL.

\section{Numerical method}
\subsection*{Numerical solution of differential equations}
There are three types of differential equations to represent physical phenomena:
\begin{itemize}
\item Time dependent equations. To solve these equations, one integrates numerically some variable in time. An example of this is the classical three body problem, where one integrates in time the position of a body in the presence of two other bodies, taking into account only their graviational interaction. Time integration uses methods like the simple but effective Euler method, or the more complicated Runge-Kutta.
\item Space dependent problems. This type of problems require integrating a differential equation independent of time, such as Laplace's equation for the electrostatic potential. When there is a spatial dependence, one resorts to methods like finite differences or finite elements. Typically an iterative method is used to find a convergent solution. 
\item Space and time dependent methods. This is the present case of TDGL equation, and it is a combination of the former two. One typically finds by spatial integration the time derivative of the variable of interest (in our case, $\frac{d\Psi(t)}{dt}$, in turn dependent on $\Psi(t)$) and integrates it in time. With the updated value of $\Psi(t)$, $d\Psi(t)/dt$ is calculated again.
\end{itemize}

\subsection*{Finite difference discretization}
Finite difference methods approximage derivatives by differences of a variable between neighboring cells of a mesh. There are three ways of approximating a derivative: forward: $\frac{dy}{dx}\simeq\frac{y(x+h)-h(x)}{h}$, central: $\frac{dy}{dx}\simeq\frac{y(x+h/2)-h(x-h/2)}{h}$ and backward: $\frac{dy}{dx}\simeq\frac{y(x)-h(x-h)}{h}$ The mesh has a constant spacing (not necessarily the same spacing in different directions). 

\subsection*{Link variable method}
We have used the usual approach to solving the TDGL equations known as link variable method. Normally in finite difference methods the interaction between neighboring cells is expressed (partially or completely) via the spatial derivatives. The interesting feature of the link variable method is that it relates the interaction between neighboring cells with the magnetic flux enclosed by each set of cells forming a closed path, which is also dependent on an external applied field. This is done by introducing the auxiliary ``link variables'' in all directions, of the form:

$$ U^x_{x,y,z}=e^{-i\int^{x}_{x_0}A_x(\xi,y,z,t)d\xi}$$
$$ U^y_{x,y,z}=e^{-i\int^{y}_{y_0}A_y(x,\eta,z,t)d\eta}$$
$$ U^z_{x,y,z}=e^{-i\int^{z}_{z_0}A_z(x,y,\zeta,t)d\zeta}$$

being $x,y,z$ the spatial components,  and $\xi, \eta, \zeta$ are auxiliary integration variables in the same directions as $x,y,z$ (not to be confused with the superconducting coherence length, $\xi$). $x_i$ are the positions where $\Psi(x,y,z)$ is calculated, and $x_0, y_0, z_0$ are arbitrary positions that in the end up cancelling out after doing line integrals along closed paths.

Given a plane, $XY$ for example, the rectangle formed by the cells in ($i,j$), ($i+1,j$), ($i,j+1$) and ($i+1,j+1$) will hold a circulation of the vector potential that can be related to magnetic flux at that position inside the superconductor via link variables:
$$ \oint _{\partial \Sigma}\vec{A} \cdot d \vec{l}=\iint_{\Sigma} \vec{H} \cdot d\vec{s}=\Phi _B$$
From such a rectangle the discretized distribution of magnetic field can be found in the $x=i$, $y=j$, $z=k$ coordinates:

$$U^x_{i,j,k}U^y_{i+1,j,k}\overline{U^x}_{i,j+1,k}\overline{U^y}_{i,j,k}=e^{-iH(i,j,k )\Delta x \Delta y}$$
and a similar calculation for the other two planes $YZ$ and $ZX$.

\subsection*{Discretization of the TDGL equations}
We start with the TDGL equations in the following dimensionless , for zero scalar potential gauge:
$$\frac{\partial\Psi}{\partial t}=\eta^{-1}[(-i\vec{\nabla}-\vec{A})^2\Psi+(1-T)(\vert\Psi\vert^2-1)\Psi]$$
$$\frac{\partial\vec{A}}{\partial t}=(1-T)Re\{\Psi^*(-i\vec{\nabla}-\vec{A})\Psi\}-\kappa^2\vec{\nabla}\times\vec{\nabla}\times\vec{A}$$

Here, lengths are scaled by the the coherence length at zero temperature $\xi_0=\xi_{T=0}$. Time $t$ is in units of $t_0 = \frac{\pi \hbar}{96 k_B T_c}$. The vector potential $\vec{A}$ is expressed in units of $H_{c2}\cdot\xi_{0}$. The coherence length dependence on temperature is assumed to be well described by $\xi(T)=\xi_{0}/\sqrt{1-T/T_c}$. $\kappa$ is the Ginzburg-Landau ratio which decides if a superconductor is type I or II. $\eta$ is a positive constant with value $\eta=\frac{\xi_0^2}{D \cdot t_0}$, with $D$ the diffusive constant for normal electrons. $Re$ means real part.\\

\begin{figure}[h]
\centering
\includegraphics[width=0.6\textwidth]{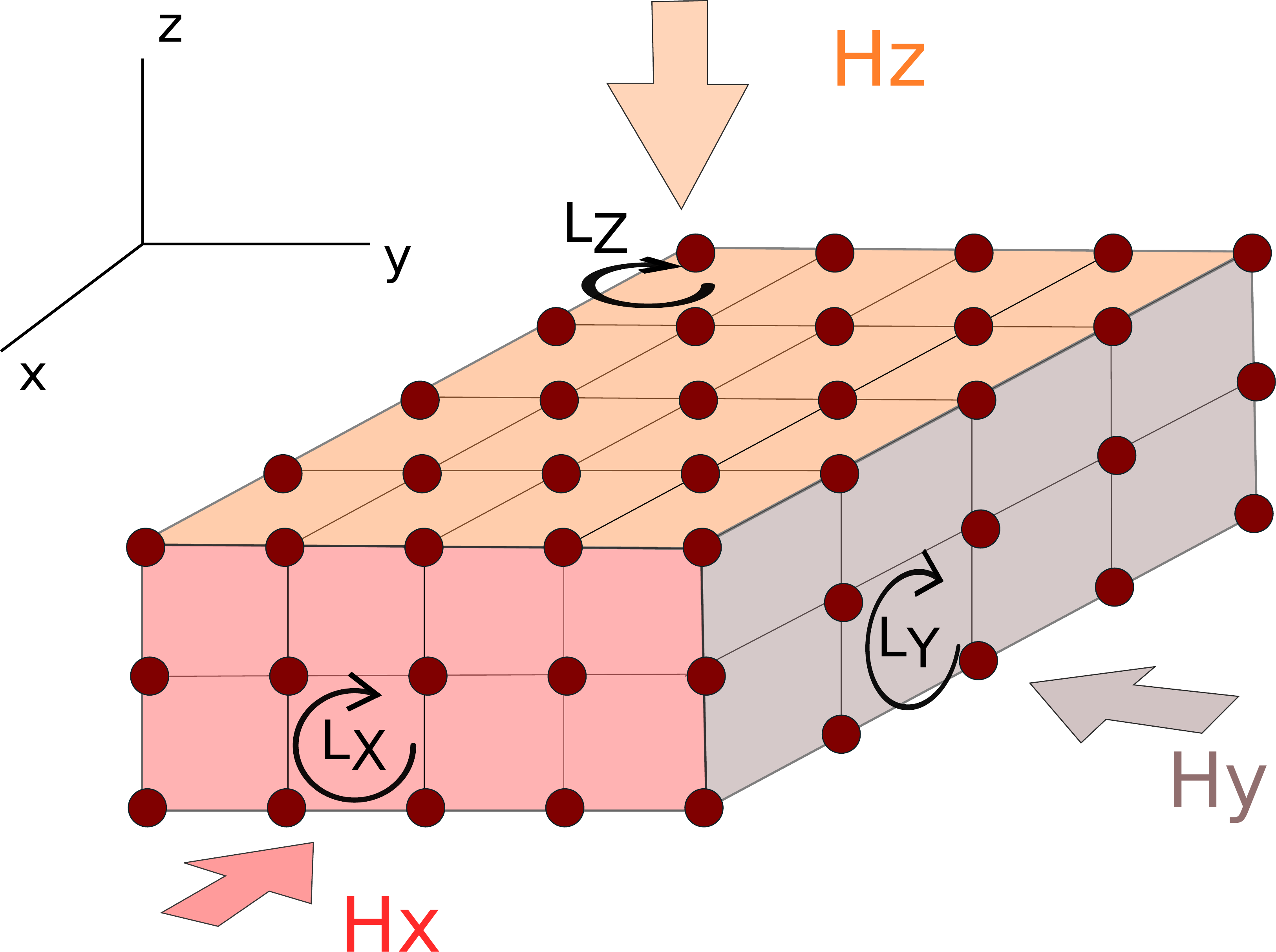}
\caption{Circulation of the link variables in a 3D simulation.}
\label{fig:LxLyLz}
\end{figure}

To reproduce the configuration of our experimental setup, we need to be able to have magnetic fields both parallel and perpendicular to the plane. With the more usual 2D configuration, one can only apply fields perpendicular to the plane, since to consider the effect of magnetic flux, it has to actually go through some closed path. If one tries to apply a field parallel to the plane ($XY$, for example) in a 2D simulation, because there is only one cell in the $Z$ direction, no flux can enter. To have nonzero magnetic flux in-plane requires having more than one cell in the perpendicular to the plane direction.
However, the more complicated geometry we want to reproduce cannot be achieved in 2D, since magnetic flux (used for boundary conditions) needs at least two layers to be accomodated in the simulated domain (flux is calculated via the circulation of link variables. Therefore, only flux perpendicular to a direction where the film has more than one layer of cells can be considered). Figure \ref{fig:LxLyLz} shows more graphically the circulation of the link variables in every direction.

Also, we need boundary conditions. The first condition is for $\Psi$ at the boundaries. We will consider the case of samples isolated from their surroundings. This means that no supercurrent flows through the boundary, and is expressed as:

$$\vec{n} \cdot (-i\vec{\nabla}-\vec{A})\Psi=0$$

A second boundary condition, for the magnetic field is used. It is through it that the magnetic field actually enters the sample. It has the form: 

$$\vec{n}_i\cdot\vec{\nabla}\times\vec{A}=H_{e,i}$$

Where $n_i$ is the unit vector in the direction of the component of the magnetic field $\vec{H}$ that we are interested in, parallel to a given surface, and $H_{e,i}$ is the value of the same component of the external magnetic field.

The equations integrated in time are four, one for the order parameter (first Ginzburg Landau equation), and the other three for the auxiliar variables known as link variables (see \cite{Buscaglia} for more details), that can be obtained by rearranging the three components of the second Ginzburg Landau equation for the vector potential. The boundary conditions, which depend on $\vec{A}$, can also be written in terms of the link variables. Therefore, the vector potential $\vec{A}$ is not explicitely solved, and it is not needed either. Its value (more precisely, its circulation) is included implicitly in the link variables, which are used to recalculate the order parameter in each step of the simulation, making it unnecesary to explicitly solve $\vec{A}$.

\begin{figure}[h]
\centering
\includegraphics[width=\textwidth]{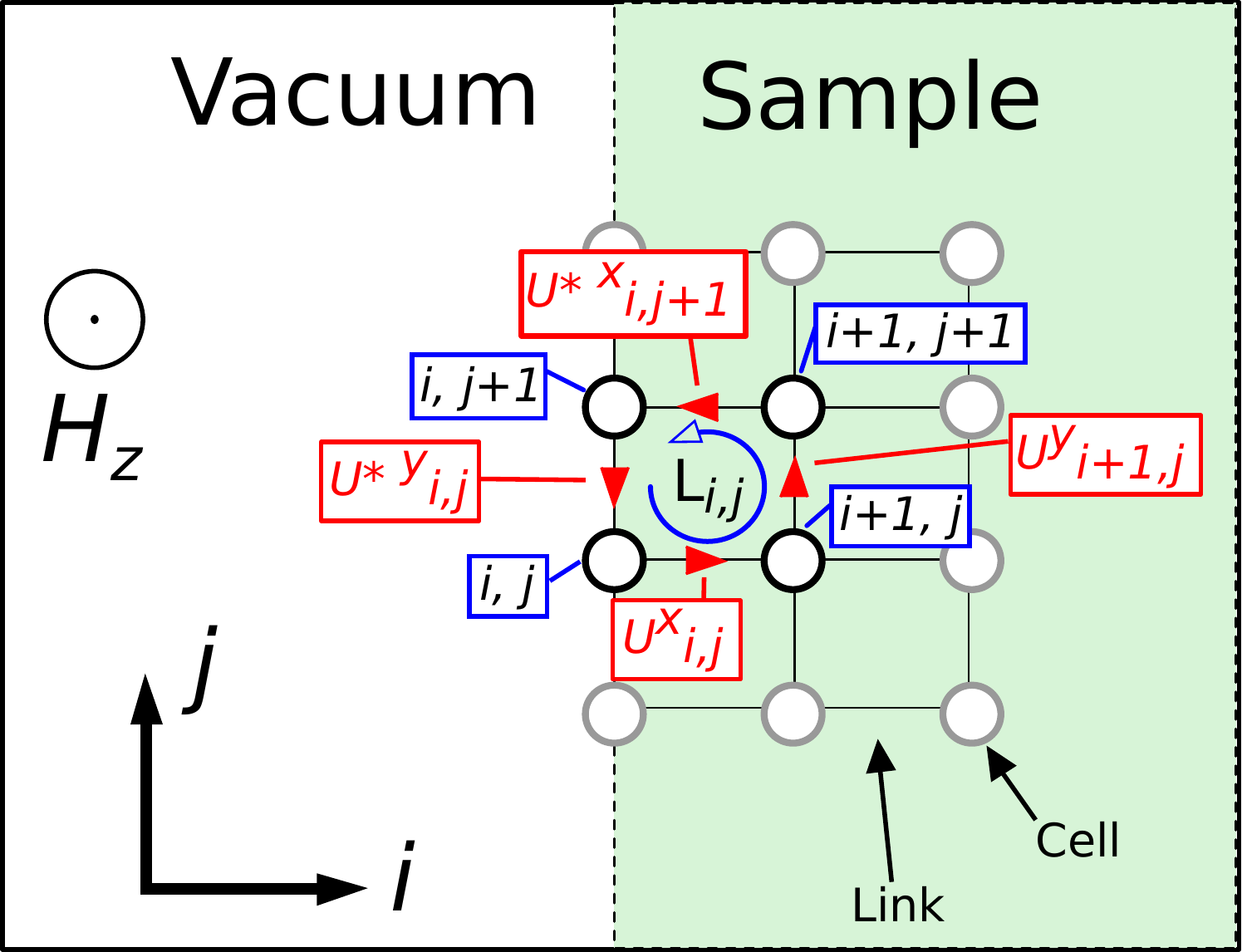}
\caption{Circulation of link variables at the boundary in 2D.}
\label{fig:boundary}
\end{figure}

An important detail encountered when transitioning from 2D to 3D is described next. To solve the equations for the link variables in the boundary in 2D one uses the circulation (see figure \ref{fig:boundary} for more clarity):
$$L_{i,j}=U^x_{i,j}U^y_{i+1,j}U^{* x}_{i,j+1}U^{* y}_{i,j}=e^{-ia_xa_yH^z_{i,j}}$$
This set of four cells is what we will refer to as ``square loop''.

By multiplying both sides by de appropriate link variable (conjugated or not) one can isolate the link variable at the edge of the sample for each of the smallest possible square loops to calculate its value from the value of the other three and the applied field. 

\subsection*{Passing from 2D to 3D}

\begin{figure}[h]
\centering
\includegraphics[width=0.5\textwidth]{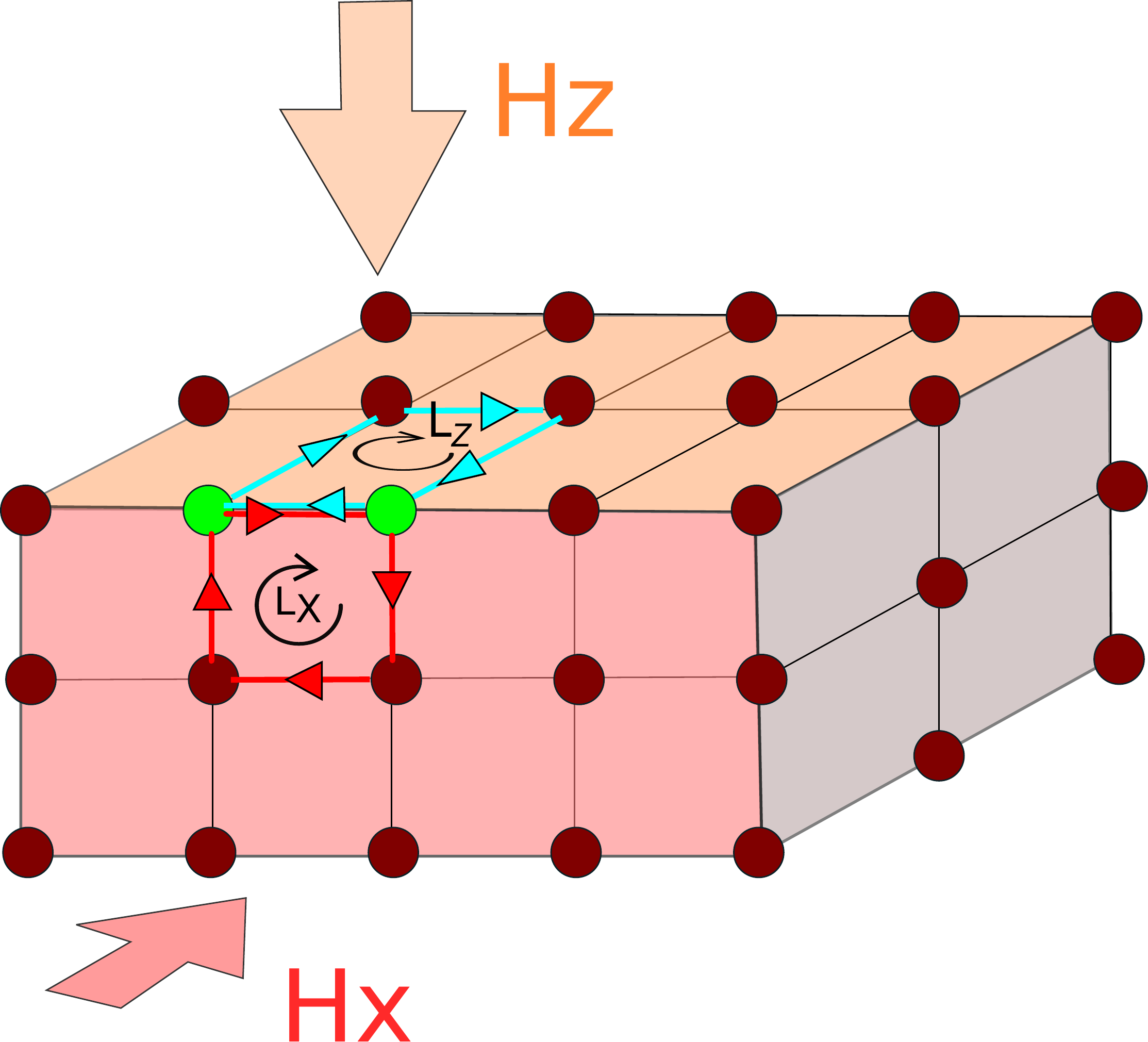}
\caption{Link variables at the edges (in this figure, between the two green cells) have an uncertain value.}
\label{fig:corner}
\end{figure}

If one tries now to do the same thing in 3D, there is an inconsistency, because at edges the unknown boundary link variable can take two different values, depending on whether it is calculated from the one or the other loop in which it takes part (figure \ref{fig:corner}). The workaround that we came up with, and works well, is to forget about calculating boundary link variables, and just force the flux through the boundary square loops. The boundary link variables are given arbitrary values. This works because these link variables really aren't used to calculate anything.

\begin{figure}[h]
\centering
\includegraphics[width=0.9\textwidth]{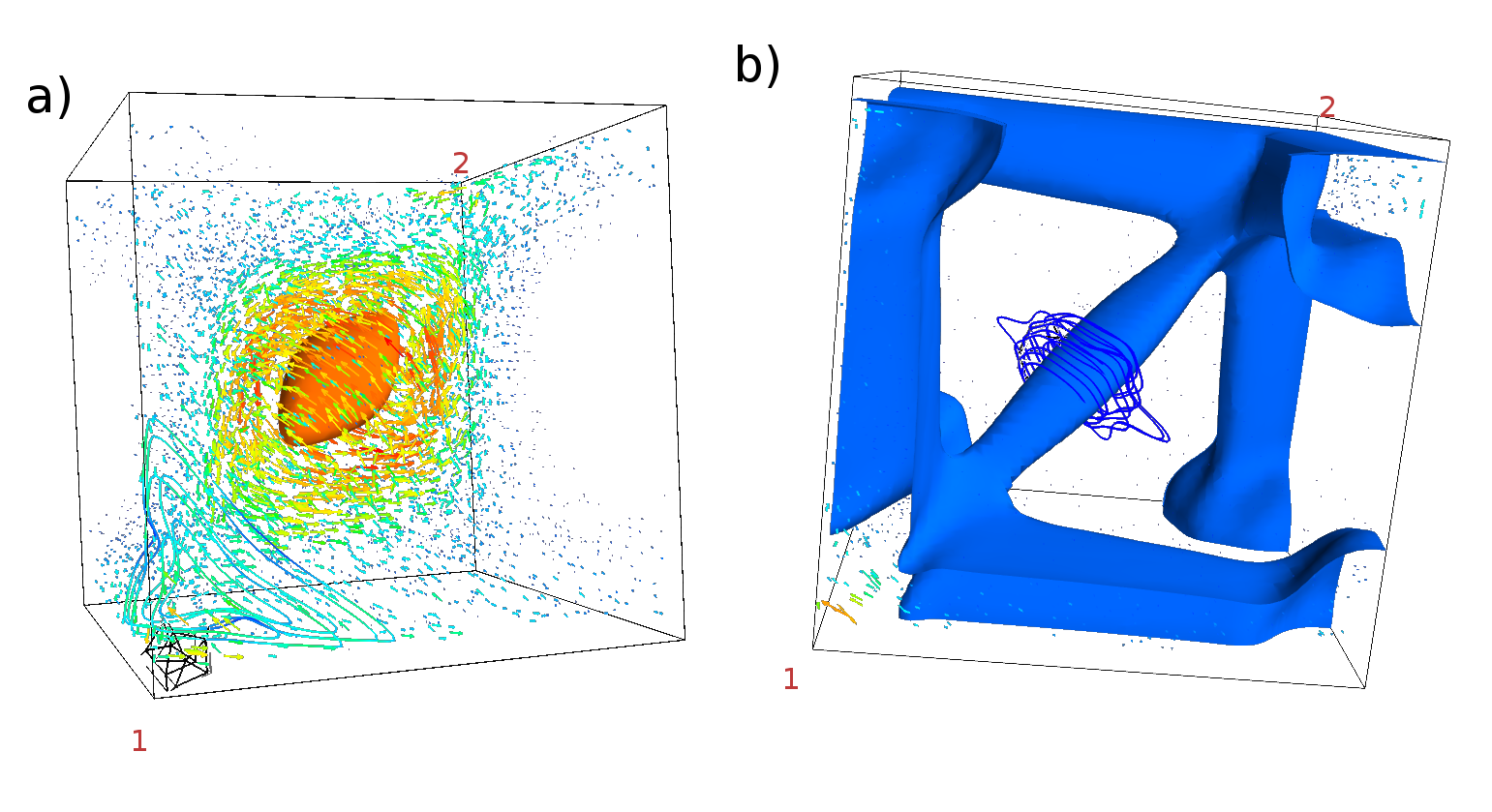}
\caption{(a) Development of a tilted vortex state at early stages and (b) isosurface for the order parameter. Field is applied in the direction (1,1,1), from corner 1 to corner 2.}
\label{fig:ball}
\end{figure}

\begin{figure}[h]
\centering
\includegraphics[width=0.9\textwidth]{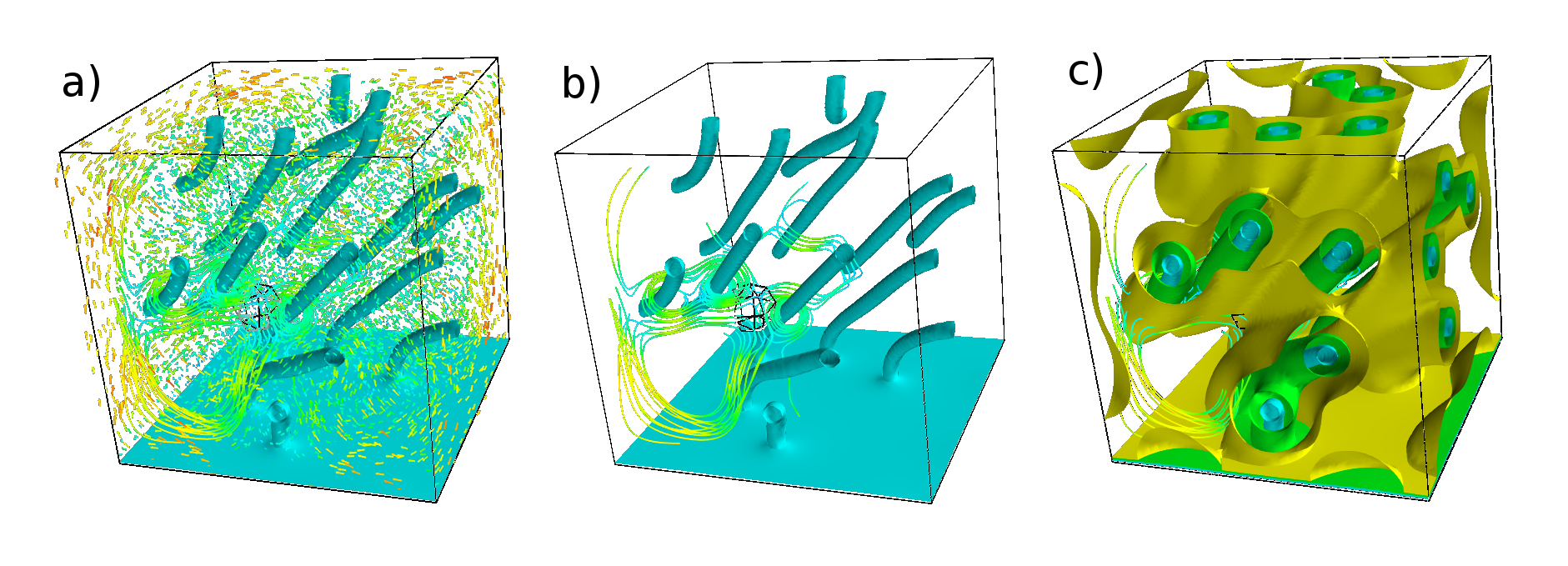}
\caption{Tilted vortex state. Isosurface at $\vert\Psi\vert=0.1$, a) with and b) without arrows representing supercurrents. c) Isosurfaces at $\vert\Psi\vert=0.1$, $0.2$ and $0.3$.}
\label{fig:tilted1}
\end{figure}

Having overcome this problem, everything works well. Then, one would expect to see tilted vortices under an inclined field, as indeed happens. Figure \ref{fig:ball} shows two isosurfaces of early stages of the development of a vortex state under an inclined field in a cube in a simulation with $\kappa=2$ and 50x50x50 cells. After some simulation time the system stabilizes and tilted vortices are observed (figure \ref{fig:tilted1}). In these scalar fields $\vert \Psi(x,y,z)\vert$ one needs to select isosurfaces at low values, for example, plot only the points with $\vert \Psi(x,y,z)\vert=0.1$ to clearly see the vortices. In this case, a magnetic field from one corner to the oposite ($(1,1,1)$ direction) is applied. Even without considering demagnetizing effects, vortices start at one border of the sample and end at another one which is not the opposite. The boundary conditions force the direction of magnetic field at the boundaries, and far from them the vortex curves to adapt. Thus, shorter vortices (closer to corners) are also more curved. Calculated isocurrents curl around vortices. Instead of plotting all (or a subset) of the arrows at each cell, as in figure \ref{fig:tilted1} a), the supercurrent direction can be easier to understand if we only plot some isolines, as in figure \ref{fig:tilted1} b). These isolines clearly show the supercurrent screening taking place in the direction perpendicular to the applied field (figure \ref{fig:ball} a), i.e., the streamlines fall in planes perpendicular to the $(1,1,1)$ direction.

\section{Results}
\subsection*{Increasing field part of hysteresis loops}

\begin{figure}
\includegraphics[width=0.8\textwidth]{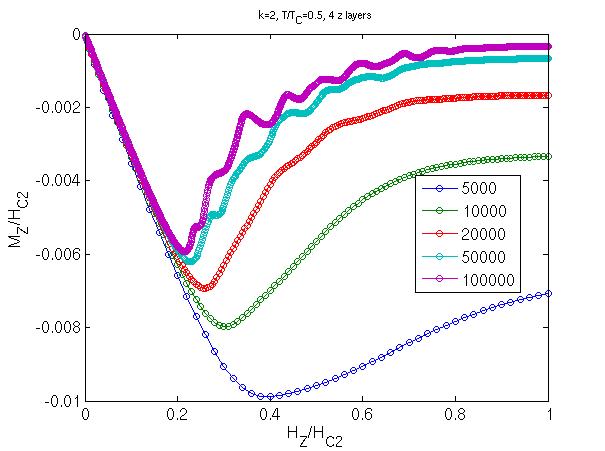}
\caption{Initial parts of hysteresis loops from $H=0$ to $H=H_{c2}$ in a different number of steps.}
\label{fig:hyst}

\end{figure}

Figure \ref{fig:hyst} shows field ramps and the global magnetization response. As expected from a superconductor, magnetization is negative for positive fields, creating a field opposing the external field, so that the internal field is zero in the superconductor. In these time dependent calculations, the speed at which the applied fields change is relevant. In the figure it is easy to see how ramping the field more slowly (more iterations) gives a curve with more jumps. This is so because vortices have more time to enter the sample discretely, overcoming at certain moments, many at once, boundary energy barriers. Faster ramps are less controlled, and vortices enter in a more disorganized manner, leading to a smoother curve, where jumps corresponding to entering vortices are smoothed out. The presence of jumps is also very dependent on the size of the sample. Their relative importance is larger in smaller samples. On the other hand, large samples give the more typical smooth ``field penetration regions'' found in magnetization measurements.

\subsection*{Steady flow of vortices between permanent magnets}
\begin{figure}[h]
\centering
\includegraphics[width=0.5\textwidth]{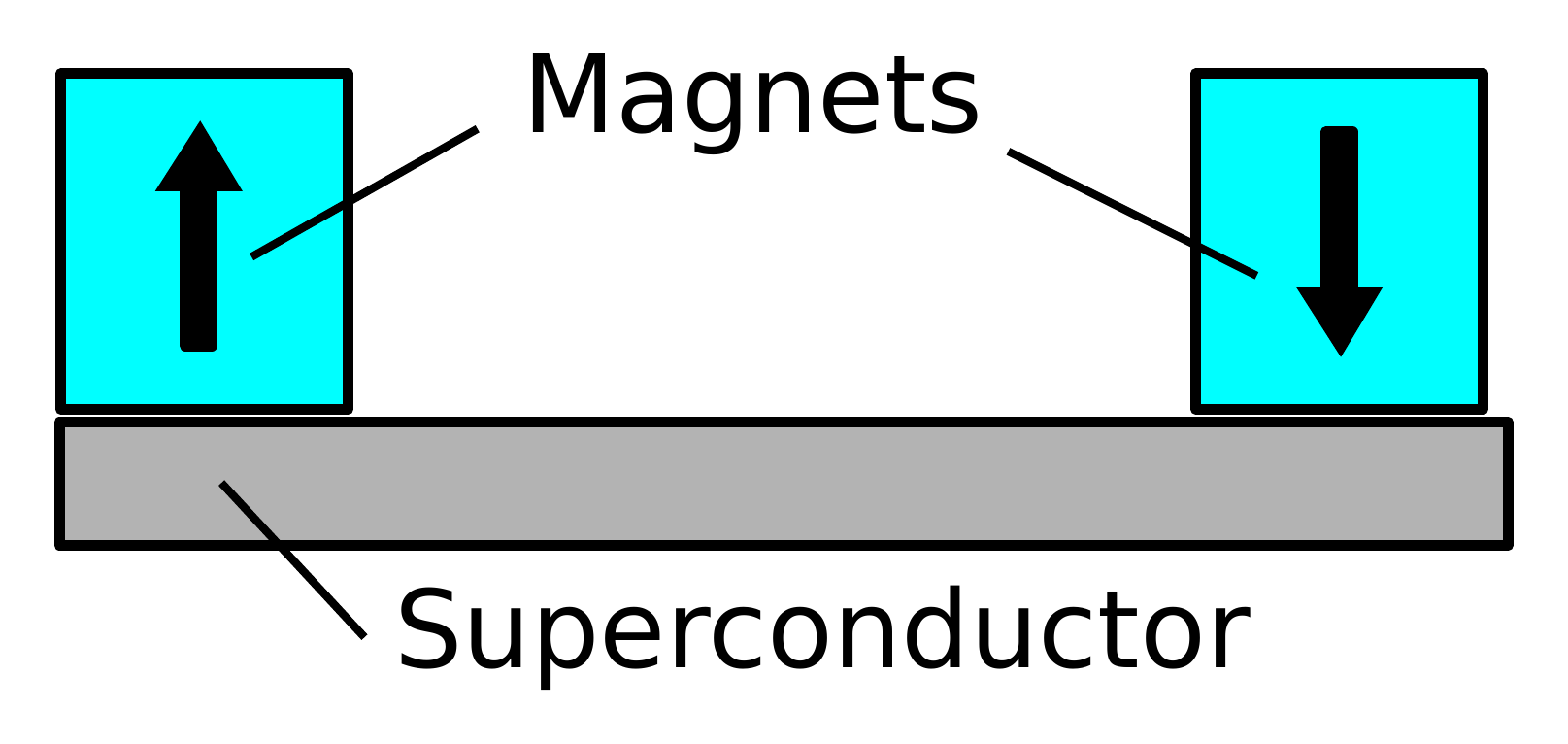}
\caption{A superconducting strip with two magnets at its ends.}\label{fig:steady}
\end{figure}

In this numerical experiment, we verify the idea of converting magnetostatic energy into a steady flow of vortices in one direction. The setup is simple: a superconducting strip, with the appropriate width to hold only one row of vortices across its length, and two permanent magnets, each one in an end of the strip, with magnetization pointing in opposite directions, perpendicular to the plane of the strip (figure \ref{fig:steady}).

The magnets are modeled by boundary conditions: a non uniform external field, nonzero at the strip ends (only in a few cells near the center, not the whole lateral side), and zero everywhere else.

\begin{figure}
\centering
\includegraphics[width=0.8\textwidth]{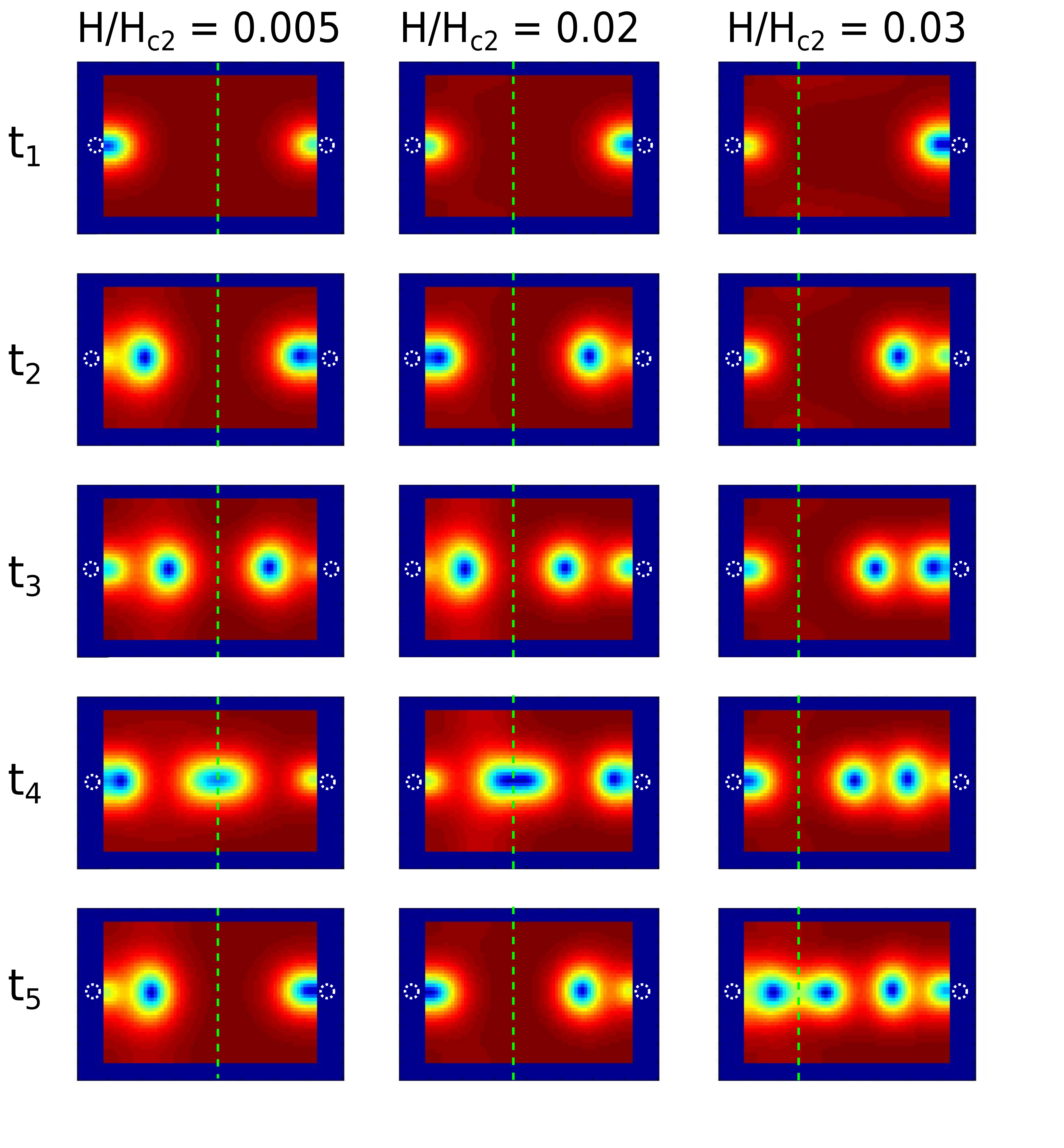}
\caption{Snapshots of the order parameter of a superconducting strip with magnets at the center of the right and left sides. The left side magnet points opposite to the external field, and the right side magnet points in the same direction as the external field. The external field is indicated at the top of each column. Rows represent snapshots at different times (evenly spaced). The green dashed line indicates the annihilation point of opposite sign vortices created at the sides. The white circles indicate the position of the magnets, which produce a magnetic field of value $H/H_{c2}=0.2$ in that region.}\label{fig:annihilation}
\end{figure}

 Of course, this is an approximation, but the fast decay of $\simeq 1/r^3$ typical of dipolar magnetic field suggests that is still valid. If both magnets are equally strong, vortices and antivortices are created in the same amount, and they meet at the center of the strip, as long as they don't need to travel a large distance (if they have, they will exit the superconductor throught the long sides). Since they have different signs, they annihilate. What can be done now is tuning the values of the strength of the magnets, making one higher than the other. In this case, the point where the vortices meet shifts laterally, closer to the weaker magnet. This is so because the stronger magnet manages to get the vortices it creates sooner into the strip than the weaker magnet. We can even get to the point where there is only a steady flow of vortices from one magnet to the other (when one of them is not strong enough to create vortices, but still stronger than the zero field at the long sides of the strip). Practically, this would mean to tune the values of very small magnets, which does not seem very feasible or practical. Instead, we can apply a perpendicular to the plane magnetic field, which will oppose the field created by one of the permanent magnets, and reinforce that of the other magnet. One needs to be careful to apply low values to this external field, since if it is too large it can lead to the creation of vortices itself. Instead, by keeping its values low, we are creating a ``slope'' which will determine where vortices of different signs will annihilate. Figure \ref{fig:annihilation} shows how the external field can influence this.

This idea could be implemented for different purposes, such as very precise magnetic field detectors, since moving vortices induce electric fields, which could be tracked by some array of sensors over the strip, to relate the annihilation position of vortices to the external field. Also, electrically charged nanoparticles could be transported by these moving vortices, until they are deposited in the annihilation place, where no longer electric field is induced after annihilation. 

\subsection*{Including holes} Another feature that could be added to the program, which fell out of the scope of this thesis due to lack of time to develop it completely, is being able to include holes in the calculations, also of arbitrary cross section but, at least to start with, ``drilled'' completely perpendicular to the boundary. Again, the idea shown in\cite{Buscaglia} could be adapted to 3 dimensions. An extra step would be necessary for including multiple holes: since for each hole, the magnetic field in that hole needs to be tracked for imposing boundary conditions, a function could scan the whole sample, to detect which cells belong to what hole. A more or less complicated procedure should be thought of to identify what is a hole, and what cells form it, based on the neighboring cells which have already been scanned and are empty. After that, the cells forming the boundary of each hole should be associated a ``type'' of cell, depending on where the neighboring empty cell lays with respect to them. With that, and knowing the field already trapped inside each hole, the boundary conditions could be calculated. Whenever a vortex falls inside a hole, the magnetic field in it will experience abrupt jumps. Including holes in the calculations would allow to simulate effects like periodic pinning, to calculate matching fields, as well as the interaction of vortices with defects of different shapes, for example, as is usually done in the context of ratchet effects with triangular defects\cite{silhanek_ratchet,vicent_triangulos}.

\subsection*{Rectification effects}
Rectification effects of the vortex lattice have attracted a lot of attention, specially in superconducting films with periodic pinning arrays, typically of triangular elements\cite{silhanek_ratchet, vicent_triangulos, vicent_rectification}. In general, nonlinear dynamics in the vortices in asymmetric potentials, under high frequency microwave or combined  DC and AC (microwave) drives has recently returned into the focus of interest of the reseachers in vortex physics and applications. \cite{Shklovskij2013,Dobrovolskiy2015,Dobrovolskiy2018}

\begin{figure}
\centering
\includegraphics[width=\textwidth]{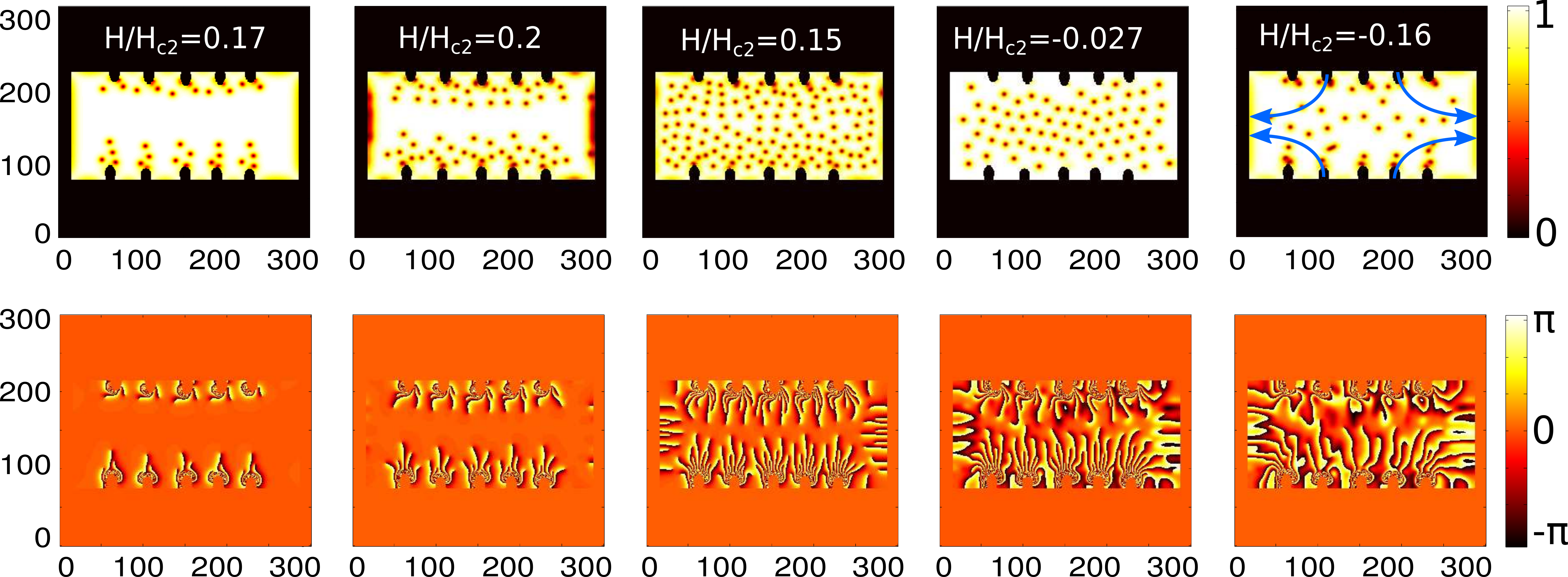}
\caption{Amplitude (top row) and phase (bottom row) of the order parameter. The perpendicular to the plane field is indicated in each column. The blue arrows indicate the net vortex flux direction.}
\label{fig:rectification}
\end{figure}

The idea of rectification is to create a potential landscape that favors the motion of vortices in one direction with respect to the opposite direction. These systems are of interest both from the applied and fundamental point of view, since they allow to design mechanisms to control the vortex flow artificially. Some other works focus on motion of vortices in structures with ratchet-like shape\cite{triangle_rectifier, rectification_MD}. Also, rectification effects have been found in systems that do not posses obvious ratchet potentials\cite{rectificador_sierra}. Those AC current induced DC voltages appear at high enough frequencies due to vortex rectification by each of the oppositely situated surface barriers\cite{Aliev2006}.  Most of the work so far considers the motion of vortices already present in the sample, whether it is a uniform film with pinning centers, or patterned superconducting elements. In this case we have applied our program to the case of vortices entering and exiting the sample. Rectification effects in this case occur due to the very geometry of the sample and the energy barriers that vortices need to overcome in order to enter inside. In this set of simulations (figure \ref{fig:rectification} we have put our program to the test to verify two facts: First, adding rectangular notches to the boundary facilitates the entrance of vortices through them. This had been already found for single triangular notches in circular samples \cite{comsol_tdgl} and one side situated triangular notches in rectangular samples \cite{Cerbu2013}. Second, a rectangular shape with the notches in the long sides, in the presence of a perpendicular to the plane field (which slowly oscillates in value), a larger number of vortices enter through these sides than they exit during an oscillation period. This means that the combination of this geometry with the notches is enough to create a net vortex flux for an external field of zero average value in time. In practice, if such mechanism proved to be true, a voltage could be measured between the sides of the sample.

\subsection*{Motion of vortices under a stationary current}
\begin{figure}
\centering
\includegraphics[width=\textwidth]{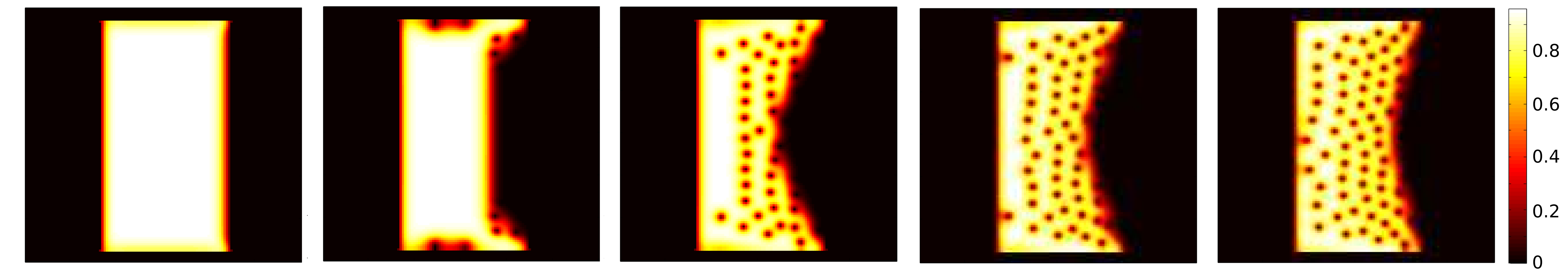}
\caption{Vortex motion perpendicular to an external current flowing along the long side of the strip (60 $\xi$ in length).}
\label{fig:flow}
\end{figure}

This simulation reproduces the vortex flow under the influece of a current flowing through a superconducting strip. A simple approximation to simulate a current is to, instead of using any current at all, modify the magnetic fields at the boundaries parallel to the flow of current. In one of them, an extra field should be added, and in the other, the same value should be subtracted, since a current flowing uniformly will create a magnetic field around it. This simplification only works for 2D simulations, due to symmetry reasons, since for 3D simulations, the curvature of the magnetic field lines should be taken into account, but in 2D they are perfectly perpendicular to the plane, since this case is equivalent to a sample infinitely long in the $z$ axis. This method has been used before \cite{IV_peeters}. Figure \ref{fig:flow} shows snapshots of a simulation of the vortex motion under a current parallel to the long sides of the strip, every 20 timesteps.

\subsection*{Spontaneous nucleation of vortices}
\begin{figure}
\centering
\includegraphics[width=\textwidth]{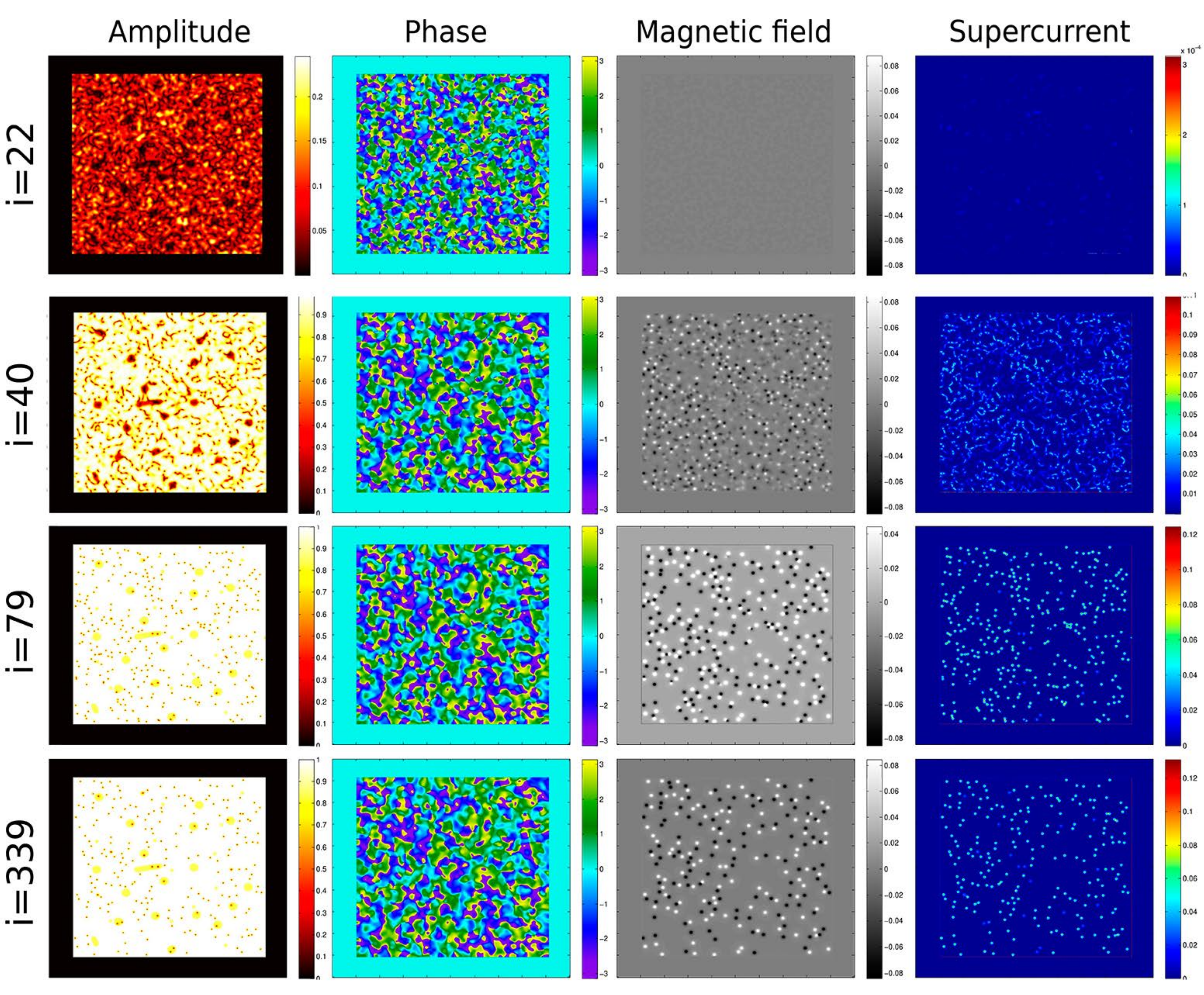}
\caption{Spontaneous vortex antivortex pairs formation starting from a random order parameter.}
\label{fig:spontaneous}
\end{figure}

Usually it is of interest to create vortices in a controlled manner with an applied field, but that is not the only way. Spontaneous formation of vortex antivortex pairs has been observed due to the stray fields of magnetic elements placed on top of superconducting films\cite{spontaneous_v_av_MFM}. A superconductor cooled from above the critical temperature can also develop pairs of vortex antivortex inside\cite{simulation_spontaneous2,simulation_spontaneous}, still having a total zero magnetic flux. This configuration is unstable, and vortices and antivortices will find it favorable to approach and annihilate each other to reduce the magnetic energy.
Next we explore the behavior of the program to calculate the spontaneous formation of vortex antivortex pairs in a superconductor in the absence of an external magnetic field, starting from a random distribution of the order parameter centered around $\vert \Psi \vert=0.1$. This is shown in figure \ref{fig:spontaneous} for a simulation with 1000 x 1000 cells (0.5$\xi$ per cell) in the plane. This replicates the situation of abruptly quenching superconductivity by cooling down a sample in the normal state. After the vortices and antivortices are formed, they start annihilating each other if they happen to be close together. After several annihilations, there will be some vortices and antivortices left, far apart enough to not feel each other's attraction, and the final state will have vortices inside, but still a zero (or close to zero, due to vortices exiting through the boundary) total magnetic field inside.

\section{Perspectives}
The numerical solution of the time dependent Ginzburg Landau equation has been an active topic since the early 1990s, until the present moment. Currently there are a number of researchers using these simulations to study theoretically the behavior of increasingly complex superconducting systems. We have focused on a rather simple system, a thin uniform film, under conditions reproducing our measurements (crossed alternating and constant magnetic fields), as well as to predict what would happen under certain conditions, paying attention to the motion of vortices. Interesting results on static 3D simulations of vortices in parallel magnetic field can be found in \cite{Wang2017}. Application of the developed method to high frequency response of the vortex system related with microwave stimulated superconductivity can be found in \cite{Lara2015,Lara2017}.
\\
The use of these simulations is far from being exhausted. Currently an important effort is being made toward understanding multiband superconductors, such as MgB$_2$, with several order parameters using numerical methods more complicated that the one presented here, where several order parameters are present and coupled among themselves\cite{multiband_milosevic}.\\
Also, as has already happened in the field of electronic circuits design, especially in circuits working at high frequencies, simulations can become a must-have tool to design better performing circuits, without the need to build lots of circuits by trial and error before reaching an optimum configuration. We expect that the same situation will be reached soon with superconducting circuits, which are gaining more and more attention for applications such as superconducting antennas and quantum computers\cite{sc_qubit}.

\section{Acknowledgements}

This work has been supported in parts by Spanish Ministerio de Ciencia, Innovación y Universidades (MAT2015-66000-P, RTI2018-095303-B-C55, EUIN2017-87474, MDM-2014-0377) and Comunidad de Madrid (NANOMAGCOST-CM P2018/NMT-4321).

\end{document}